\documentclass[11pt,a4paper]{article}
\usepackage{delarray,amsfonts,amsmath,amssymb}
\usepackage{graphicx}
\DeclareGraphicsExtensions{ps,eps,ps.gz}

\renewcommand\doteq{\equiv}

\makeatletter
\def\@citex[#1]#2{\if@filesw\immediate\write\@auxout
        {\string\citation{#2}}\fi
\def\@citea{}\@cite{\@for\@citeb:=#2\do
        {\@citea\def\@citea{,}\@ifundefined
        {b@\@citeb}{{\bf ?}\@warning
        {Citation `\@citeb' on page \thepage \space undefined}}
        {\csname b@\@citeb\endcsname}}}{#1}}
\newif\if@cghi
\def\cite{\@cghitrue\@ifnextchar [{\@tempswatrue
        \@citex}{\@tempswafalse\@citex[]}}
\def\citelow{\@cghifalse\@ifnextchar
[{\@tempswatrue\@citex}{\@tempswafalse\@citex[]}}
\def\@cite#1#2{{\if@cghi\unskip$\null^{#1}$\else #1\fi\if@tempswa\typeout
        {warning: optional citation argument ignored: `#2'} \fi}}

\begin{document}

\def\rr{{\mathbb{R}}}
\def\oo{{\cal O}}
\def\sh{\text{ sh}\,}
\def\ch{\text{ ch}\,}
\def\th{\text{ th}\,}
\newcommand{\norm}[1]{\left\lVert#1\right\rVert}
\newcommand{\abs}[1]{\lvert#1\rvert}
\def\aa{{\cal A}}
\def\rrr{{\cal R}}
\def\mm{{M}}
\newcommand{\scal}[1]{\langle#1,#1\rangle}
\newcommand{\scl}[2]{\langle#1,#2\rangle}
\def\dm{\lp\begin{array}}
\def\fm{\end{array}\rp}
\def\ee{{\cal E}}
\def\lb{\left[}
\def\rb{\right]}
\def\lp{\left(}
\def\rp{\right)}
\newcommand{\tr}[1]{\text{Tr}(#1)}
\def\ttt{{\cal T}}
\def\nnn{{\cal{N}}}

\title{\bf {\Large Diamonds's Temperature:} \\
{\large   Unruh effect for bounded trajectories
 and thermal time hypothesis
}}

\author{Pierre Martinetti ${}^{ab}$, Carlo Rovelli ${}^{ac}$\\[.2cm]
${}^{a}${\em Centre de Physique Th\'eorique de Luminy, }
{\em Case 907, F-13288, Marseille}\\
${}^{b}${\em Laboratoire de Physique Th\'eorique
Universit\'e Mohammed 1, M-60000, Oujda}\\
${}^{c}$ {\em Dipartimento di Fisica Universit\`a La Sapienza, PA
Moro 2, I-00185, Roma}\\
{\em martinet@cpt.univ-mrs.fr, rovelli@cpt.univ-mrs.fr}}

\date{\small\today} \maketitle

\begin{abstract}

We study the Unruh effect for an observer with a finite lifetime,
using the thermal time hypothesis.  The thermal time hypothesis
maintains that: (i) time is the physical quantity determined by
the flow defined by a state over an observable algebra, and (ii)
when this flow is proportional to a geometric flow in spacetime,
temperature is the ratio between flow parameter and proper time.
An eternal accelerated Unruh observer has access to the local
algebra associated to a Rindler wedge. The flow defined by the
Minkowski vacuum of a field theory over this algebra is
proportional to a flow in spacetime and the associated temperature
is the Unruh temperature.  An observer with a finite lifetime has
access to the local observable algebra associated to a finite
spacetime region called a ``diamond".  The flow defined by the
Minkowski vacuum of a (four dimensional, conformally invariant)
quantum field theory over this algebra is also proportional to a
flow in spacetime.  The associated temperature generalizes the
Unruh temperature to finite lifetime observers.

Furthermore, this temperature does not vanish even in the limit in
which the acceleration is zero.  The temperature associated to an
inertial observer with lifetime $\cal T$, which we denote as
``diamond's temperature", is $\ T_{D}=2\hbar/\pi k_{b}{\cal T}$.
This temperature is related to the fact that a finite lifetime
observer does not have access to all the degrees of freedom of the
quantum field theory. However, we do not attempt to provide any
physical interpretation of our proposed assignment of a
temperature.

\end{abstract}

\section{Introduction}

This work has two motivations.  First, to study the Unruh
effect\cite{unruh} for an observer with a finite lifetime. Second,
to probe the thermal time hypothesis\cite{carloconnes}. The Unruh
effect is the theoretical observation that the vacuum state
$\Omega$ of a quantum field theory on Minkowski spacetime looks
like a thermal equilibrium state for an uniformly accelerated
observer with acceleration $a$.  The observed temperature is the
Davis-Unruh temperature $T_{U}=\hbar a/2\pi k_{b}c$, and is
closely related to Hawking's black hole temperature ($c$ is the
speed of light, $\hbar$ and $k_{b}$ are the Planck and Boltzmann
constants.)

There exist many derivations of the Unruh effect.\cite{Wald,pauri}
One generally assumes that the observer moves over along an
infinite worldine of constant acceleration.  A model detector
moving on this world line and interacting with the vacuum state of
the field becomes excited and reaches a thermal state.  For a free
field, this can be computed integrating the two-point functions
along the infinite trajectory.  Alternatively, one can consider
the region causally connected to the world line.  This region is
the Rindler wedge $W$.  A quantization scheme for the field
restricted to $W$ leads to the definition of particle states.  The
conventional vacuum $\Omega$ is a thermal distribution state of
these particle.  Another approach is based on Bisognano-Wichman
theorem\cite{bisognano,sewell,haag} : World lines of uniformly
accelerated observers in $W$, parametrized by proper time, are
orbits of the action of a one-parameter group, which can be
interpreted as the time flow of the accelerated observers.  The
vacuum state $\Omega$ is a KMS state with temperature $T_{U}$,
with respect to this time flow, over the algebra of the local
observables in $W$.  Intuitively, the reason the accelerated
observer does not see a pure state can be viewed as related to the
fact that he has no access to part of $\Omega$.  Each observer
instantaneity surface is cut into two parts by the edge of the
wedge.  The degrees of freedom of the field on the other side of
the edge are inaccessible to the observer.  Since $\Omega$ has
vacuum correlations across the edge, the two sets of degrees of
freedom on the two sides of the edge are entangled.  The
restriction of $\Omega$ to the edge interior, therefore, fails to
be pure a state.

Doubts have been raised on the universality of the Unruh effect, for
instance in
Ref.[\citelow{fedotov}].  One of the sources of the doubts is that most
derivations of the effect consider infinite trajectories, while the
trajectories
of realistic detectors have a beginning and an end.  In our opinion, the
theoretical support for the Unruh effect is rather robust, and these doubts
do
not have much consistency.  Nonetheless, the issue of observers with finite
lifetime is interesting.  The region causally connected to a finite
trajectory
is far smaller than $W$.  Typically, it is the intersection of the future of
the
beginning point of the trajectory with the past of its end point.  Such a
region
is called a ``diamond".  In suitable Lorentz coordinates $x^\mu=(x^0,\vec
x)$, any
diamond can be represented as the region $D_L$, with $L>0$, defined by
\begin{equation}
\abs{x^0} + \abs{\vec x} < L.
\end{equation}
${\cal T}=2L$ is the lifetime of the observer if he moves inertially.
In a sense, only the finite region $D_L$ can take part in the
thermalization process. More precisely, the observer has access
only to the degrees of freedom of the field that lie within
$\abs{\vec x} < L$ on the $x^0=0$ simultaneity surface.  Does this
observer see a thermal state?  Is there a sense in which there is
an associated temperature?  These questions can be addressed in a
variety of ways.  Here we
consider the possibility of adapting the Bisognano-Wichman
approach to this case.

Let us now come to the thermal time hypothesis.  Consider a system with a
large
number of degrees of freedom.  We describe our experimental knowledge of its
state as a statistical state.  A generic state determines a flow: In quantum
theory, a state over an observable algebra determines a flow called the
modular
flow, or the Tomita flow.  In the classical case, the relation between a
state
and a its modular flow reduces to the relation between the distribution
$\rho$
on phase space representing the state and the Hamiltonian flow of $\
\ln\rho$.
The thermal time hypothesis demands that the flow determined by the
statistical
state coincides with what we perceive as the physical flow of time.  A second
part of the hypothesis refers to the special cases in which a geometrical
background provides an independent notion of temporal flow: if the geometric
flow and the modular are related, the ratio of the rates of the two flows is
identified as the temperature.  The hypothesis was initially motivated by the
problem of time in quantum gravity.

The Unruh effect can be interpreted as an example in which the thermal time
hypothesis is realized.  The relation between the Unruh effect and the
thermal
time hypothesis is the following.  The vacuum state $\Omega$, restricted to
the
algebra of the observables in the wedge $W$ generates a modular flow.  This
modular flow is precisely proportional to the proper time flow of uniformly
accelerated observers.  For each observer, the ratio of the two flows is
$T_{U}$.  According to the thermal time hypothesis, this means that the
natural
physical time generated by $\Omega$ on the algebra of the observables in the
wedge is the proper time of an accelerated observer, and the corresponding
temperature is the Davis-Unruh temperature.

A natural question, then, is whether there are other cases in which the
thermal
time associated to a spacetime region has a physical meaning.  Local algebras
associated to diamonds are very natural objects in algebraic quantum field
theory.  As mentioned, a diamond is the region causally connected to an
observer
with finite lifetime.  It is natural to wonder whether the modular flow
generated by the vacuum restricted to the local algebra associated to a
diamond
--the ``diamonds' flow", or the ``diamonds' time"-- may have any
significance at all.

In general, there is no reason to expect that a modular flow should have a
geometric interpretation and lead to a notion of temperature.  Remarkably,
however, the modular flow of a diamond \emph{is} associated to a geometric
action in the case of a conformally invariant quantum field theory in four
dimensions.  Does this lead to a notion of temperature?  The modular
parameter
$s$ cannot be directly proportional to the proper time along the geometric
flow,
because the modular parameter runs all along $\rr$ while the proper time
$\tau$
of an observer in $D_L$ is bounded.  However, we can nevertheless define a
local
temperature as the \emph{local} ratio of the two flows.  This allows us to
define a temperature which is a natural generalization of the Unruh-Davies
temperature to observers with a finite lifetime. In this paper we explicitly
compute this temperature.  For fixed acceleration,
this temperature reduces to the Unruh temperature when $L$ is large.

Furthermore, this temperature does not vanish even for non-accelerating
observers.  The (minimal) temperature associated an inertial observer with
lifetime $\cal T$ is the temperature
\begin{equation}
T_{D}=\frac{2\hbar}{\pi k_{b} {\cal T}}. \label{td}
\end{equation}
This temperature is related to the fact that a finite lifetime
observer does not have access to all the degrees of freedom of the
field.  But we leave any physical interpretation of this
temperature for future investigations. In particular, it would be
interesting to study whether it is related to the response of
model detectors.

Section 2 briefly presents our main tools (definition of local observable
algebra, modular automorphism and KMS condition) and the thermal time
hypothesis.  In Section 3, we recall the discussion of the Unruh effect in
the
Bisognano-Wichman approach.  Section 4 presents our new results, namely the
discussion of the generalization of the Unruh effect to diamonds $D_L$ and
its
physical interpretation.  Section 5 briefly discusses the case of the future
cone.  Minkowski metric has signature $(+,-,-,-)$. From now on, we take
$c=\hbar=k_{b}=1$.

\section{Tools}

\subsection{Local observable algebra}

Consider a quantum field theory on four dimensional Minkowski space.  We
restrict our attention to conformally invariant quantum field theories.  The
conformal group is the group of transformations $x\mapsto x'$ that preserve
the
causal structure: $(x'{}^{\mu}-y'{}^{\mu})(x'_\mu - y'_\mu) = 0$ iff
$(x^{\mu}-y^{\mu})(x_\mu - y_\mu) = 0$.  In four dimensions, this is the
fifteen-parameters group generated by Poincar\'e transformations, dilatations
and proper conformal transformations.  An infinitesimal transformation is
given
by
\begin{equation}
\delta x^\mu= \omega^\mu{}_{\nu}x^\nu +a^\mu+\lambda x^\mu+ |x|^2 k^\mu
-2x^\mu k_\nu x^\nu,
\end{equation}
where $(\omega_{\mu\nu}=- \omega_{\nu\mu},a^\mu,\lambda,k^\mu)$ are
infinitesimal
parameters.  We assume that both the dynamics and the vacuum of our quantum
field are invariant under these conformal transformations.  Since mass terms
break conformal invariance, this means that we are dealing with a massless
field.  Equivalently, we are considering a regime in which masses can be
neglected.

We recall a few elements on local observable algebras.\cite{haag,wightman}\
\ \
We have a Hilbert space $H$ carrying a unitary representation $U$ of the
Poincar\'e group. If the theory is conformally invariant, $U$
extends to a representation of the entire
conformal group.  There is a ray $\Omega$ in $H$, called the physical vacuum,
invariant under the action of the Poincar\'e group (or the conformal group).
A field is an operator
valued distribution over Minkowski space $\mm$.  There is no operator
corresponding to the value $\phi(x)$ of the field $\phi$ at a given point
$x$;
there is an operator $\phi(f)$ corresponding to the quantity obtained
smearing
out the field with a smooth function $f$
\begin{equation}
\phi(f) = \int \phi(x)\, f(x)\ d^4 x.
\end{equation}
Consider a finitely extended open subset $\oo$ of $\mm$.  The
algebra generated by all operators $\phi(f)$, where $f$ has
support in $\oo$, is interpreted as representing physical
operations that can be performed within $\oo$.  By completing this
algebra in the norm topology, we obtain a $C^*$-algebra
$\rrr(\oo)$.  This is called the local observable algebra
associated to $\oo$. We have an algebra $\rrr(\oo)$ associated to
each open subset $\oo$.  The Reeh-Schlieder theorem states that
the vacuum $\Omega$ is a cyclic and separating vector for
$\rrr(\oo)$ as soon as the causal complement of $\oo$ (the set of
all points which lie space-like to all points of $\oo$) contains a
non-void open set.\footnote{$\Omega$ is cyclic but not separating
for the algebra associated to the entire Minkowski space $M$.
Intuitively, a separating state can be thought as a density matrix
with nonvanishing components on \emph{all} basis states.}

\subsection{Modular automorphisms}

The Tomita-Takesaki theorem associates a preferred flow $\sigma_s$ to any
$C^*$-algebra $\aa$ acting on a Hilbert space with a preferred cyclic and
separating vector $\Omega$.  The flow is built as follows.  Let $S$ be the
conjugate linear operator from $\aa \Omega$ to $\aa\Omega$
\begin{equation}
Sa\Omega \doteq a^*\Omega
\end{equation}
for any $a\in \aa$. $S$ is closeable with respect to the strong
topology and its closure \-- still denoted by $S$\-- has a unique
polar decomposition
\begin{equation}
S = J\Delta^{1/2}
\end{equation}
where $\Delta$ is a selfadjoint positive operator (unbounded in general)
called
modular operator and $J$ is antiunitary.  This allows us to define a
one-parameter group of automorphisms of $\aa$: the modular group $\sigma_s$
of
$\aa$ associated to $\Omega$
\begin{equation}
\label{sigmat}
\sigma_s\, a \doteq \Delta^{is} a \Delta^{-is}
\end{equation}
(see [\citelow{tomtak}] for details on the definition of $\Delta^{is}$) with
$s\in\rr$.  If $\aa$ is the algebra of local observables on an open set
$\oo$ of
Minkowski space, with non-void causal complement, we call {\it modular group
of
$\oo$} the modular group of $\rrr(\oo)$ associated to the vacuum $\Omega$.
We
denote it as $\sigma_{s}(\oo)$. We write $\Delta(\oo)$ for the corresponding
modular operator.

\subsection{KMS  condition}

Let us now recall the relation between modular flow and thermodynamics.
Consider a physical system with a finite number of degrees of freedom, say
$N$
particles in a finite box with volume $V$.  In statistical physics, an
equilibrium state is a state of maximal entropy.  If $H$ is the hamiltonian
describing $S$, the mean value of any observable $a$ when $S$ is in an
equilibrium state at inverse temperature $\beta$ is
\begin{equation}
\label{eq}
\omega(a) = \frac{1}{Z}\text{Tr}\lp{e^{-\beta H}a}\rp
\end{equation}
where the partition function $Z$ is given by
\begin{equation}
\label{Z} Z\doteq \tr{e^{-\beta H}}.
\end{equation}
Consider now the thermodynamical limit of this system: send the volume of the
box $V$ and the number of particles $N$ to infinity, keeping the density
$N/V$
fixed.  In this limit the total energy $E$ goes to infinity and therefore the
partition function (\ref{Z}) is no longer defined.  One therefore needs a
way to
characterize the equilibrium states alternative to equations (\ref{eq}) and
(\ref{Z}).  This can be obtained as follows.  An equilibrium state $\omega$
of
inverse temperature $\beta$ can be directly characterized as a state over the
algebra satisfying the KMS condition.  This requires that for any two
observables $a$ and $b$
\begin{equation}
\label{kms1}
 \omega( (\alpha_t a) b ) = \omega(b(\alpha_{t+i\beta} a))
 \end{equation}
where $\alpha_t$ denotes the time translation
\begin{equation}\alpha_t a = e^{iHt}ae^{-iHt}
\end{equation}
extended to a complex argument, and that the function
\begin{equation}
\label{kms2} z\mapsto \omega(b(\alpha_z a))
\end{equation}
is analytic in the strip
\begin{equation}
0 < \text{Im}\, z < \beta.
\end{equation}
A state $\omega$ satisfying these conditions is called KMS with respect to
the
time translation $\alpha_t$.  One can shows that both conditions (\ref{kms1})
and (\ref{kms2}) reduce to (\ref{eq}) in the case of a finite dimensional
system.  These conditions represent the correct extension of the definition
of
an equilibrium state to infinite-dimensional systems.\cite{haag}

Now, the remarkable fact is that, given a $C^*$-algebra $\aa$ acting on a
Hilbert space $H$ with a cyclic and separating vector $\Omega$,
the faithful state $\omega$ over $\aa$ defined by
\begin{equation}
\label{vide}
\omega(a) = \scl{\Omega}{a\Omega}
\end{equation}
for any $a\in\aa$ (we use Dirac notations) is KMS with respect
to the modular group $\sigma_s$ defined by $\Omega$. More
precisely one has
\begin{equation}\omega((\sigma_s a)b) = \omega(b(\sigma_{s-i}a)).
\end{equation}
Define $\alpha_{\beta s} \doteq \sigma_s$. Then
\begin{equation}\omega((\alpha_{-\beta s} a)b) = \omega(b(\alpha_{-\beta s
+ i\beta}a)).
\end{equation}
Therefore, if we \emph{define}
\begin{equation}
\label{temps}
 t\doteq -\beta s
\end{equation}
we obtain condition (\ref{kms1}) (as far as (\ref{kms2}) is
concerned, see ref.~[\citelow{haag}]).  In other words: {\it an
equilibrium state at inverse temperature $\beta$ can be
characterized as a faithful state over the observable algebra
whose modular automorphism group $\sigma_s$ is the time
translation group.  The modular parameter $s$ is proportional to
the time $t$.  The proportionality factor is minus the inverse
temperature.} (The minus sign is there only for historical
reasons: the mathematicians have defined the modular flow with
sign opposite to that given by the  physicists).

\subsection{Thermal time hypothesis}

The modular group equips a system in a generic state with a dynamics.  The
existence of such an intrinsic dynamics suggests a solution to a longstanding
open puzzle in the physics of general covariant systems.  The dynamical laws
of
a generally covariant system determine correlations between observables, but
they do not single out any of these observables as time.  This ``timeless"
structure is sufficient to provide all dynamical predictions. However, this
timeless structure leaves
us without any understanding of the physical basis for the flow of time,
a flow that has evident
observable effects particularly in thermodynamical contexts.  The
{\it
thermal time hypothesis} is the idea that there is no dynamical basis for the
flow of time, but there is a thermodynamical basis for it.  More precisely, a
generic state determines a modular flow, and it is this flow that represents
the
physical basis of the flow of time.  This flow is state dependent.  Thus:

\noindent {\it The physical time depends on the state.  When the system is
in a
state $\omega$, the physical time flow is given by the modular flow of
$\omega$}.

The second part of the hypothesis demands then that if the modular flow is
proportional to a flow in spacetime, parametrized with the proper time
$\tau$,
then the ratio of the two flows is the temperature
\begin{equation}
\beta=\frac{1}{T}\doteq -\frac{\tau}{s}.
\label{beta}
\end{equation}
In other words, when $t$ is independently defined, (\ref{temps}) becomes a
definition of the temperature.  In previous applications of this idea, only
cases in which $\frac{\tau}{s}$ is exactly constant along the flow were
considered.  In this paper, we extend the framework slightly, by considering
a
case in which the modular flow is proportional to a flow in spacetime, but
the
constant of proportionality between $\frac{d}{ds}$ and $\frac{d}{d\tau}$
varies
along the flow.  In such a situation, it is natural to consider a local
notion
of temperature, defined at each point by
\begin{equation}
\beta(s)=\frac{1}{T(s)}=-\frac{d\tau(s)}{ds}.
\label{tth2}
\end{equation}

The thermal time hypothesis is based on the idea that the origin of
time is thermodynamical, and that the time flow is state-dependent.  In a
sense, the thermal time hypothesis is an inversion and a generalization of KMS
theory. Given a time flow $\sigma$, KMS theory identifies the thermal states
as the KMS states of the time flow.  But given the KMS state, the time flow
$\sigma$ is precisely its modular flow. Since any generic
state defines a modular flow, the idea here is that, generically, this modular flow is
in fact the physical time that governs macroscopic thermodynamics.

As far as conventional systems are concerned, notice the following remarkable
fact: if we measure the statistical distribution of an equilibrium state, we
can
infer the hamiltonian from the result.  We can then define time as the
parameter
of the flow generated by the hamiltonian.  This provide an operational
characterization of time.  On the other hand, it is well known that
it is difficult to provide a purely
dynamical (as opposed to statistical/thermodynamical) characterization of
what
is time.  In a generally covariant context, where definitely no physical
time is
determined by the dynamics, the thermal time hypothesis identifies the
physical
time as the modular flow for generic states.

The thermal time hypothesis has been checked in several examples.  The most
striking one concerns the cosmic microwave background (CMB). The statistical
state distribution describing the CMB determines a flow that coincides with
the
time coordinate of usual Friedman-Robertson-Walker metric.\cite{carlo93bis}\
\
The associated temperature is the CMB temperature.  The Rindler wedge
provides
another example in which the hypothesis is realized.

For a detailed discussion of this issue and the motivations of the thermal
time
hypothesis, see refs.~[\citelow{carloconnes},\citelow{carlo93}].  On the
large
literature on problem of time in general relativity see for instance
ref.~[\citelow{carlo91}] and references therein.

\section{Time in the Rindler Wedge}

Consider a uniformly accelerated observer.  For simplicity, let us
assume that the motion is in a plane with constant $x^2$ and
constant $x^3$, and that the acceleration four-vector writes $(0,
a, 0, 0)$ in the local observer frame. Here $a\in \rr^+$.  The
world line of this observer can be written as\cite{landau}
\begin{equation}
\label{universeline}
 x^1 = \sqrt{\frac{1}{a^2} + (x^0 + C')^2}+ C
 \end{equation}
where $C$ and $C'$ are constants. With a Poincar\'e transformation, we can
take $C$ and $C'$ to zero
\begin{equation}
 x^1 = \sqrt{a^{-2} + (x^0)^2}.
\label{ao}
 \end{equation}
A Lorentz boost in the $x_1$ direction reads
\begin{equation} \Lambda(\rho) = \dm{cccc} ch\, \rho & sh\, \rho & 0 & 0 \\
sh\, \rho& ch\, \rho & 0 & 0 \\ 0 & 0 & 1 & 0 \\ 0&0&0&1 \fm
\end{equation}
($\rho\in \rr$).  The trajectory (\ref{ao}) is an orbit of this group.  The
proper time $\tau$ along the orbit is proportional to the boost parameter
$\rho$, and the proportionality factor is given by $a$
\begin{equation}
\rho = a \tau .
\end{equation}
We can therefore parametrize the trajectory (\ref{ao}) with the proper time
$\tau$ and write it as
\begin{equation}
 x^\mu(\tau) = (a^{-1} \text{sh}\, a\tau, \
a^{-1} \text{ch}\, a\tau,\ 0,\ 0)
\end{equation}
 so that
\begin{equation}
\Lambda(a\tau) x(\tau_0) = x(\tau_0 + \tau).
\end{equation}
That is, $\Lambda(a\tau)$ generates a proper-time translation of proper time
$\tau$ along the accelerated world line.

Consider a quantum field theory defined on $M$ and let $K$ denote the
representation of the generator of this boost, so that
\begin{equation}U(\Lambda(\rho)) = e^{i\rho K}.\end{equation}
Then the operator
\begin{equation}
U(\tau)=e^{i a\tau K}
\label{tempsgeo}
\end{equation}
can be viewed as an operator that generates the evolution in
proper-time seen by the accelerated observer.

The region causally connected to the accelerated observer, namely the set of
the
points that can exchange signals with a uniformly accelerated observer is the
Rindler wedge $W$, defined by
\begin{equation}
\label{wedge} x^1 > \abs{x^0}.
\end{equation}
The wedge $W$ has a non void causal complement thus the modular
group $\sigma_s(W)$ is defined. The modular operator $\Delta(W)$
is known\cite{bisognano}. It corresponds precisely to a Lorentz
boost
\begin{equation}
\Delta(W) = e^{-2\pi K}.
\end{equation}
therefore the modular flow $\sigma_s(W)$ (defined in
(\ref{sigmat})) is given by the boost
\begin{equation}
U(s) = e^{-2\pi s iK}.
\end{equation}
This flow is precisely proportional to the geometric flow (\ref{tempsgeo}).
The
relation between the modular parameter $s$ and the proper time $\tau$ is
obtained comparing the two operators: a translation by $s$ generates a shift
in
proper time
\begin{equation}
\tau = -\frac{2\pi}{a} s.
\end{equation}
From (\ref{beta}), the temperature is determined by
\begin{equation}
{\beta} = -\frac{\tau}{s} = \frac{2\pi}{a}.
\end{equation}
The result is therefore that the state $\Omega$, restricted to the Rindler
wedge
generates a time flow which is the one of an accelerated observer, with an
associated temperature which is the Unruh-Davis temperature.
\begin{equation}
T =\frac{1}{\beta} = \frac{a}{2\pi} = T_{U}.
\end{equation}

\section{The time of the diamonds}

\subsection{Diamond's modular group}

Consider now an observer that lives for a finite time.  Let $x_{i}$ be the
event
of his birth and $x_{f}$ the event of his death, which is in the past of
$x_{i}$.  To begin with, assume that the life of the observer is the straight
segment from $x_{i}$ to $x_{f}$.  Let
\begin{equation}
{\cal T}= 2L = |(x_{f}-x_{i})|
\end{equation}
be the observer lifetime, namely the proper time between $x_{i}$ and
$x_{f}$.
Let $D$ be the region which is in the past of $x_{f}$ and in the future of
$x_{i}$.  This region is called a Diamond.  It is the region with which the
observer can exchange signals during his lifetime: he can send a signal, and
receive a response, to each point of $D$.  Imagine that the observer sends
out
measuring apparata on spacecrafts to make local measurements on the quantum
field, and receives back the signals from his apparata.  Then he will be
capable
of observing, at most, the local observables in $D$.  The local algebra to
which
he has access is therefore $\rrr(D)$.  Assuming that the quantum field
theory is
in its vacuum state $\Omega$, we can ask what kind of state will this
observer
observe.  The answer is the restriction of $\Omega$ to $\rrr(D)$.  Let us
therefore study the properties of this restriction.

With a Poincar\'e transformation, we can always choose Minkowski coordinates
such that
\begin{equation}
 x_{i}^\mu=(-L,0,0,0), \hspace{3em}
 x_{f}^\mu=(L,0,0,0).
\end{equation}
The region $D$ is then the region defined by
\begin{equation}
\abs{x^0} + \abs{\vec x} < L.
\end{equation}
Now, the key observation is that there is a conformal transformation
$K:M\mapsto M$ that sends the diamond $D$ to the wedge $W$
\begin{equation}
K(D) = W.
\end{equation}
Explicitly, if $L=1$, $K$ is given by
\begin{eqnarray}
 x'{}^\mu=\frac{2x^\mu+\delta_{1}^\mu(1+\abs{x}^2-2x^1)}{1-\abs{x}^2-2x^1}.
\end{eqnarray}
Using this map, it is immediate to find out the modular flow
$\sigma_s(D)$ of the diamond local algebra $\rrr(D)$: this is
simply obtained by mapping the modular flow of the wedge to the
diamond\cite{hislop}. More precisely, we consider the one
parameter group of transformations $\Lambda_{D}(\rho)$ of the
diamond into itself given by
\begin{equation}
\Lambda_{D}(\rho)\doteq K^{-1}\Lambda(\rho) K.
\end{equation}
Since this is a composition of conformal transformations,
$\Lambda_{D}$ is a one-parameter subgroup of conformal
transformations.  A straightforward calculation (see ref.
\citelow{haag}) gives $\Lambda_{D}(\rho): x_\mu\mapsto
x_\mu(\rho)$ where
\begin{equation}
x^\mu(\rho) = \frac{2x^\mu +\delta^\mu_{0} \lp 2x^0 \text{ ch}
\,\rho + (1 + x^\mu x_\mu) \text{sh}\, \rho -2x^0\rp}{2 x^0 \text{
sh}\,\rho +(1 + x^\mu x_\mu)\text{ ch} \,\rho + (1 - x^\mu x_\mu)}
.
\end{equation}
As in the case of the wedge, the modular operator $\Delta(D)^{is}$ is the
representation of the conformal transformation $\Lambda_{D}(\rho)$ for $\rho=
-2\pi s$.

This generalizes easily to a diamond $D_L$ of arbitrary dimension
$L > 0$. The modular operator $\Delta(D_L)^{is}$ is the
representation of the conformal transformation
$\Lambda_{D_{L}}(\rho)={\cal L}\Lambda_{D}(\rho){\cal L}^{-1}$
where ${\cal L}$ is the dilatation by a factor $L$. Explicitly
\begin{equation}
x^\mu(\rho) =L\ \frac{2Lx^\mu +\delta^\mu_{0} \left( 2Lx^0\
\text{ch} \,\rho + (L^2 + x^\mu x_\mu)\, \text{sh}\, \rho
-2Lx^0\right)}{ 2 Lx^0\ \text{sh}\,\rho +(L^2 + x^\mu x_\mu)
\text{ ch} \,\rho + (L^2 - x^\mu x_\mu)}. \label{TL}
\end{equation}

\subsection{Temperature}

The orbits of the modular group of the diamond are uniformly accelerated
trajectories.  Indeed, consider a uniformly accelerated observer moving from
$x_{i}$ to $x_{f}$ with acceleration $a$.  For simplicity, assume that the
motion is in the plane $x^2=x^3=0$.  Its worldline is given by
(\ref{universeline}) with $C'=0$ and
\begin{equation}
\label{C}
C= -a^{-1}\sqrt{1
+a^2L^2}.
\end{equation}
We can parametrize this orbit using the proper time along the orbit. This
gives
\begin{equation}
\label{tempspropre} x^\mu(\tau)  = (a^{-1} \text{sh}\, a\tau,\
a^{-1} \text{ch}\, a\tau+C,\ 0,\ 0)
\end{equation}
where $\tau$ runs
from $-\tau_0$ to $\tau_0$ with
\begin{equation}\tau_0\doteq a^{-1}\text{ arcsh }a L.\end{equation}
The line of universe (\ref{tempspropre}) is precisely an orbit of
the diamond's modular group $\sigma_s(D_L)$. Let us fix the origin
of the modular parameter $s$ so that $s=0$ when $\tau=0$. This
gives $x^\mu=x^\mu(0)=(0,a^{-1}+C,0,0)$. Inserting this value of
$x^\mu$ in (\ref{TL}) gives, with simple algebra,
\begin{eqnarray}
\label{xsmu}
x^0(\rho) &=& \frac{L a^{-1}\text{ sh}\,\rho}{a^{-1}\text{ ch}\,\rho -C},\\
\nonumber x^1(\rho) &=& \frac{L^2}{C- a^{-1}\text{ ch}\,\rho}
\end{eqnarray}
and $x^2(\rho) = x^3(\rho) = 0$. Indeed, it is not difficult to
check that for any $\rho\in \rr$, $x_1(\rho) -C$ is positive and
satisfies $(x_1(\rho) -C)^2 = a^{-2} + x_0(\rho)^2$ so
$(\ref{xsmu})$ is solution of (\ref{ao}). Moreover
\begin{equation}
\lim_{\rho\rightarrow -\infty} x(\rho) = x_{i}, \hspace{2em}
\lim_{\rho\rightarrow \infty} x(\rho) = x_{f}
\end{equation}
thus the orbit of $x(\rho)$ under the action of $\sigma_s(D_L)$
coincides with the worldline of the uniformly accelerated
observer.

We have therefore two parameterizations of the same worldline.
One (\ref{tempspropre}) given by the proper time
$\tau\in ]-\tau_0, \tau_0[$, the other (\ref{xsmu}) given by the
modular parameter $\rho=-2\pi s\in ]-\infty, +\infty[$. To find the
relation between these two parameterizations, we can compare
$x^\mu(\tau)$ with $x^\mu(\rho)$. This gives
\begin{eqnarray}
x^0\ \ =& a^{-1}\ \text{sh}\, a\tau &=\ \ \frac{L a^{-1}\text{
sh}\,\rho}{a^{-1}\text{ ch}\, \rho -C},\\
x^1\ \ =& a^{-1}\ \text{ch}\, a\tau +C &=\ \ \frac{L^2}{C- a^{-1}\text{ ch}\,
\rho}.
  \end{eqnarray}
It is convenient to write $\lambda\doteq \text{ arcsh}\, a L$, so that
$aC= -\text{ ch}\,\lambda$. Using this, we have
\begin{equation}\sh a\tau = \frac{\sh \rho \sh \lambda}{\ch \rho + \ch
\lambda}\;,\quad \ch a\tau = \frac{1 + \ch \lambda \ch \rho}{\ch \rho +
\ch \lambda}.
\end{equation}
This gives
\begin{equation}
\label{taudet} \tau(\rho) = \frac{1}{a}\text{arcth}\left(
\frac{\sh \rho \sh \lambda}{1 + \ch \lambda \ch \rho} \right).
\end{equation}
Differentiating, we get
\begin{equation} \frac{d\tau(\rho)}{d\rho} = \frac{L}{\ch \rho + \ch
\lambda}.
\end{equation}
Recalling that $\rho=-2\pi s$, and the definition of $\lambda$ we conclude
\begin{equation}
\frac{d\tau(s)}{ds} = -2\pi \frac{d\tau(\rho)}{d\rho} =
-\frac{2\pi L}{\ch \rho + \ch \lambda}= -\frac{2\pi
L}{\sqrt{1+a^2L^2}+\ch(2\pi s)}.
\end{equation}

The thermal time hypothesis (\ref{tth2}) defines the temperature
\begin{equation}
\beta(s) = -\frac{d\tau(s)}{ds} =  \frac{2\pi
L}{\sqrt{1+a^2L^2}+\ch(2\pi s)}. \label{temp}
\end{equation}
Noticing from (\ref{xsmu}) that $\beta=-\frac{2\pi}{La}x^1$,
we can use (\ref{tempspropre}) to rewrite this inverse temperature as
\begin{equation}
\label{temptau}
\beta(\tau)  = \frac{2\pi}{La^2} \
(\sqrt{1+a^2L^2}-\text{ch}\, a\tau)
\end{equation}
In particular, at $x^0=0$, that is $s=0$, we have
\begin{equation}
\beta_{0} \doteq \beta(0) =
 \frac{2\pi L}{\sqrt{a^2L^2+1}+1}.
 \label{exact}
\end{equation}
We analyze the meaning of this result in the following two sections.

\subsection{Unruh effect for the observer with a finite lifetime}

First, consider the case in which $a$ is large. This corresponds
to orbits that stay near the boundary of the diamond. In
particular, at $x^0=0$ the observer is near the edge $\abs{\vec
x}=L$ of the diamond. In the limit of large $a$, we have
\begin{equation}
\beta_{0} = \frac{2\pi}{a},
\end{equation}
which is the Davis-Unruh temperature.  Therefore, the thermal time hypothesis
associates precisely the Davis-Unruh temperature to a uniformly accelerated
observer \emph{with a finite lifetime} that passes near the edge. For fixed
acceleration $a$, an observer with finite lifetime, in the middle of his
life,
observes the Davis-Unruh temperature if his lifetime $2L$ is large enough.
The
first order correction in $1/L$ gives
\begin{equation}
\beta_{0} = \frac{2\pi}{a}-\frac{2\pi}{a^2L},
\end{equation}
while the exact formula for every $L$ is given by (\ref{exact}).
This result provides a complete generalization of the Unruh effect
to observers with finite lifetime.  In order to better appreciate
this result, consider the dependence of the temperature from the
proper time. In the limit of large $L$, this is given by
\begin{equation}
\beta(\tau) =
\frac{2\pi}{a}\left(1-\frac{\ch(a\tau)}{aL}\right).
\end{equation}
In Figure 1 we have plotted this function in the finite region of the
observer
proper time.  Notice that the function goes rapidly to zero at the
boundaries of
the interval of the life of the observer, but it is nearly flat over most of
the
region.  That is, for the observer measuring his proper time, $\beta(\tau)$
is
constant and $\sim \beta_{0}$ for most of his lifetime.

\begin{figure}[ht]
\begin{center}
\mbox{\rotatebox{0}{\scalebox{1}{\includegraphics{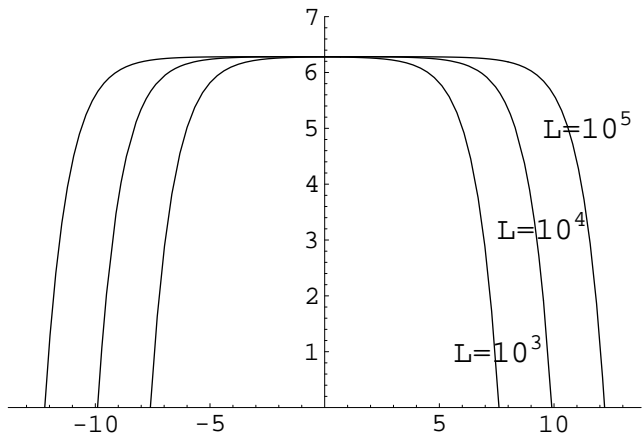}}}}
\end{center}
\caption{The inverse temperature ${\beta}$ as a function of the
proper time $\tau$ seen by the accelerated observer with finite,
but long, lifetime $L$ (acceleration is $a = 1$).}
\end{figure}

\subsection{Diamond's temperature}

Next consider the case in which, for a given lifetime $\cal T$, the
acceleration is
small.  In particular, consider the case in which $a=0$.  This is the
observer
moving along a straight line from $x_{i}$ to $x_{f}$.  Interestingly, the
temperature associated to this observer does not vanish.  The inverse
temperature $\beta$ reaches a maximum value at $x^0=0$, which is
\begin{equation}
\beta_{D} =  \pi L.
\end{equation}
Therefore the thermal time hypothesis associates the finite temperature
\begin{equation}
T_{D} = \frac{1}{\beta_{D}}=  \frac{2}{\pi {\cal T}}
\end{equation}
to an observer with a finite lifetime $\cal T$.  Inserting back the dimensional
constants $s,\hbar$ and $k_{b}$, we get equation (\ref{td}).  What is the
meaning
of this temperature?

It is tempting to conjecture that this temperature is due to the
fact that the observer cannot observe all the degrees of freedom
of the quantum field.  In particular, he can only observe the
degrees of freedom that at time zero are inside the sphere
$\abs{\vec x}<L$.  Because of the quantum field theoretical
correlations between the inside and the outside of this sphere,
the vacuum state restricted to this sphere is not a pure state. If
this is correct, there should be a physical temperature naturally
associated to observers with a finite lifetime, whether they are
accelerating or not. We leave the study of this idea for future
investigations (comparison with entanglement entropy of
[\citelow{culetu}] may be interesting).

\section{Future cone}

For completeness, we describe here also an observer that is born
at $x_{i}$ and lives forever.  The region causally connected to
this observer is the future cone $V^+$ of $x_{i}$.  Since this has
a non void causal complement, the vacuum restricted to $V^+$ is a
separating state and we can search the action of its modular group
$\sigma_s(V^+)$.  In fact this action is also geometrical. Taking
$x_{i}$ in the origin, it corresponds to the dilatation\cite{haag}
\begin{equation}
\label{paramT} x^{\mu} \mapsto x^\mu(s) \doteq e^{-2\pi s}
x^{\mu}.
\end{equation}
As for the wedge and the diamond, we can interpret the orbits of this action
as
motions of a physical observer.  An inertial observer
with constant speed $\vec v=(v^1,v^2,v^3)$ with respect to the
origin, which leaves the origin at $\tau=0$ follows the world line
\begin{equation}
\label{paramtau} x^\mu(\tau) = K^\mu \tau
\end{equation}
where $K^0\doteq c\gamma$, $K^i = v^i\gamma$ with $\gamma\doteq
(1-\frac{v^2}{c^2})^{-\frac{1}{2}}$.  The (closure of the) line of universe
(\ref{paramT}) for $s\in\rr$ coincides with the line of universe
(\ref{paramtau}) where $\tau$ runs from $0$ to $+\infty$.  The relation
between
the modular parameter $s$ and the proper time $\tau$ is  easily
obtained as
\begin{equation}
\tau(s) = e^{-2\pi s }.
\end{equation}
Therefore the temperature is
\begin{equation}
\beta(s)= 2\pi e^{-2\pi s}.
\end{equation}
The temperature is finite and non zero at the birth of the
observer and then it goes exponentially to zero, as it should for
a long living observer moving with constant speed.

\section{Conclusion}

Starting from the thermal time hypothesis, we have obtained the result that
the  Unruh effect exists also for observers with a finite lifetime.
Restauring dimensional constants, the thermal time hypothesis associates
the temperature (\ref{temptau})
\begin{equation}
T(\tau)  = \frac{\hbar La^2}{2\pi k_{b}c^3\
\left(\sqrt{1+\frac{a^2L^2}{c^4}}-\text{ch}\,
\frac{a\tau}{c}\right)}
\end{equation}
to an observer living for a finite time and moving with constant finite
acceleration $a$ between two spacetime points at proper temporal distance
${\cal T}=2L/c$ from each other.
The temperature depends on the observer proper time $\tau$
lapsed from the middle of his lifetime.  In the limit
of large $L$, we recover the Davis-Unruh temperature.
For large  $L$, an observer with a large acceleration sees the Davis-Unruh
temperature for most of his lifetime.

Furthermore, it follows from the thermal time hypothesis that also an
inertial
observer with finite lifetime has an associated temperature.  The minimum
value
of this temperature is the diamonds temperature $T_{D}$ (\ref{td})
\begin{equation}
T_{D}=\frac{2\hbar}{\pi k_{b} {\cal T}}.
\end{equation}
 We have speculated that this temperature reflects the fact that the observer
has no
access to distant degrees of freedom.  It would be interesting to understand
whether this temperature is related to the response of model thermometers
with
finite lifetime.

Altogether, the thermal time hypothesis seems to play an interesting role in
these physical situations.  Much more work is needed to study
the validity of this
hypothesis, and its relation with thermodynamical and statistical
temporal phenomena.

\subsection*{Acknowledgments}
\noindent Thank to G. Poggiospola for its helps in drawing with
Mathematica. P. Martinetti was supported by an postdoctoral AUF
grant.

\end{document}